\documentclass[preprint,secnumarabic, amssymb, nobibnotes, aps, prb]{revtex4-2}

\usepackage{graphicx}
\usepackage{bm}
\usepackage{amsmath}
\usepackage{tcolorbox}

\begin{document}

\title{Second-harmonic generation from singular metasurfaces}%

\author{Fan Yang}%
\email{fyang@scu.edu.cn}
\affiliation{College of Physics, Key Laboratory of High Energy Density Physics and Technology of the Ministry of Education, Sichuan University, Chengdu, Sichuan 610065, China}
\author{Cristian Cirac\`i}%
\affiliation{Istituto Italiano di Tecnologia, Center for Biomolecular Nanotechnologies, Via Barsanti 14, 73010 Arnesano, Italy}

\begin{abstract}

We present a theoretical study of second-harmonic generation from a singular metasurface. The singular metasurfaces strongly interact with the incident light, where the large field enhancement forms an intense surface polarization that generates the second-harmonic field. By using transformation optics, the calculation of nonlinear optical response is converted from the metasurface frame to that of a simple slab geometry, largely reducing the complexity of the problem. In addition, the singular metasurface exhibits a weak dependence on the incident angle of light, which can be potentially used as an all-angle device for harmonic generations. Finally, we study the symmetry dependence of second-harmonic generation in the far field for the singular metasurface and show how to enhance the conversion efficiency under normal incidence by breaking the surface inversion symmetry.    

\end{abstract}

\maketitle

\section{Introduction}

Optical nonlinear effects \cite{boyd2020nonlinear, shen1984principles} have received considerable interest since the discovery of second-harmonic generation (SHG) in 1961 \cite{franken1961generation}.
With the development of nanofabrication processes in the past decades, nanoscale structures have become an important component in nonlinear optics. Unlike traditional nonlinear crystals, nanostructured materials can provide strong light-matter interaction and simultaneously relax the phase-matching condition constraint \cite{celebrano2015mode, noor2020mode}.
Nanostructures may be classified into two groups: all-dielectric and plasmonic systems \cite{kauranen2012nonlinear, kivshar2018all, krasnok2018nonlinear}. All-dielectric nanostructures utilize both electric and magnetic resonances, whose excitation can contribute to an efficient harmonic generation \cite{shcherbakov2014enhanced, kruk2015enhanced}. In contrast, plasmonic structures take advantage of the excitation of surface plasmon polaritons, creating a strong field enhancement at the metal surface. This strong field then gives rise to a large nonlinear optical response \cite{kim2008high} in a close by dielectric \cite{nielsen2017giant} or in the metal itself \cite{czaplicki2018less} through the dynamics of free-electron \cite{krasavin2018free, khalid2020enhancing}. Hybrid metal-dielectric resonators have also been proposed \cite{shibanuma2017efficient}.

Many scenarios have been suggested to further enhance the nonlinear response of nanostructures. For instance, the application of epsilon-near-zero (ENZ) concept boosts the nonlinear response \cite{reshef2019nonlinear, neira2015eliminating, alam2016large}. Boyd \textit{et al.} have shown that ITO films display ENZ properties at near-infrared frequencies, leading to a giant Kerr nonlinearity \cite{alam2016large}. The concept of bound state in the continuum (BIC) \cite{hsu2016bound} has also been employed to greatly enhance SHG because the excitation of BIC contributes to a large quality factor \cite{carletti2018giant, koshelev2020subwavelength}. 
Another method to further enhance the nonlinear effect is introducing a singularity in the nanostructure. A singularity can be an ultra-small gap between two metallic interfaces or a sharp metallic tip \cite{pendry2012transformation}. These singular structures exhibit a much stronger field enhancement than conventional plasmonic structures, so a giant nonlinear response is expected near the singularities. The nonlinearity enhancement by singularities has been confirmed in a variety of experiments, such as nanoparticle dimers\cite{slablab2012second}, bowtie antennas\cite{kim2008high}, particles on metal surface\cite{li2021light}, etc.

Even though nonlinear nanophotonics has received substantial attention for a few decades, most research relies on experimental measurements and numerical simulations. Only a few works give analytic solutions to the nonlinear optical response of nanostructures, such as flat surface, cylinder, and sphere \cite{sipe1980analysis, dadap1999second, capretti2014full}.        

In the past decade, transformation optics has been successfully applied in the field of plasmonics, providing the tools to convert calculations for a complex nanostructure into a simple slab geometry problem\cite{pendry2012transformation, aubry2013transformation}. After obtaining the analytical solution in the slab frame, it is in fact possible to derive the corresponding field profile of the nanostructure by mapping the fields between two frames. Recently, this powerful analytical tool has been utilized to study the SHG from a kissing nanowire dimer \cite{reddy2019surface, elkabetz2021optimization}. However, a transformation optics approach to extended structures such as a metasurface has not been explored yet. 

Based on our previous work about the linear optical response of singular metasurfaces \cite{yang2018transformation}, in this article, we analytically investigate the SHG process in analogous systems. In particular, we consider the nonlinear response of free electrons at the metal surface \cite{ciraci2012origin, ciraci2012second}. We show that the intense field enhancement in the singular metasurface provides a strong nonlinear  excitation for the second-harmonic field. In addition, the singular metasurface supports a hidden dimension that enables a continuous mode excitation when changing the k-vector of incident waves \cite{pendry2017compacted}, thereby leading to an all-angle high SHG efficiency. Finally, the singular metasurface possesses a few surface symmetries that strongly affect SHG efficiency in the far field.        

\section{Mapping of nonlinearity}

\begin{figure}[ht]
\includegraphics[width=0.9\columnwidth]{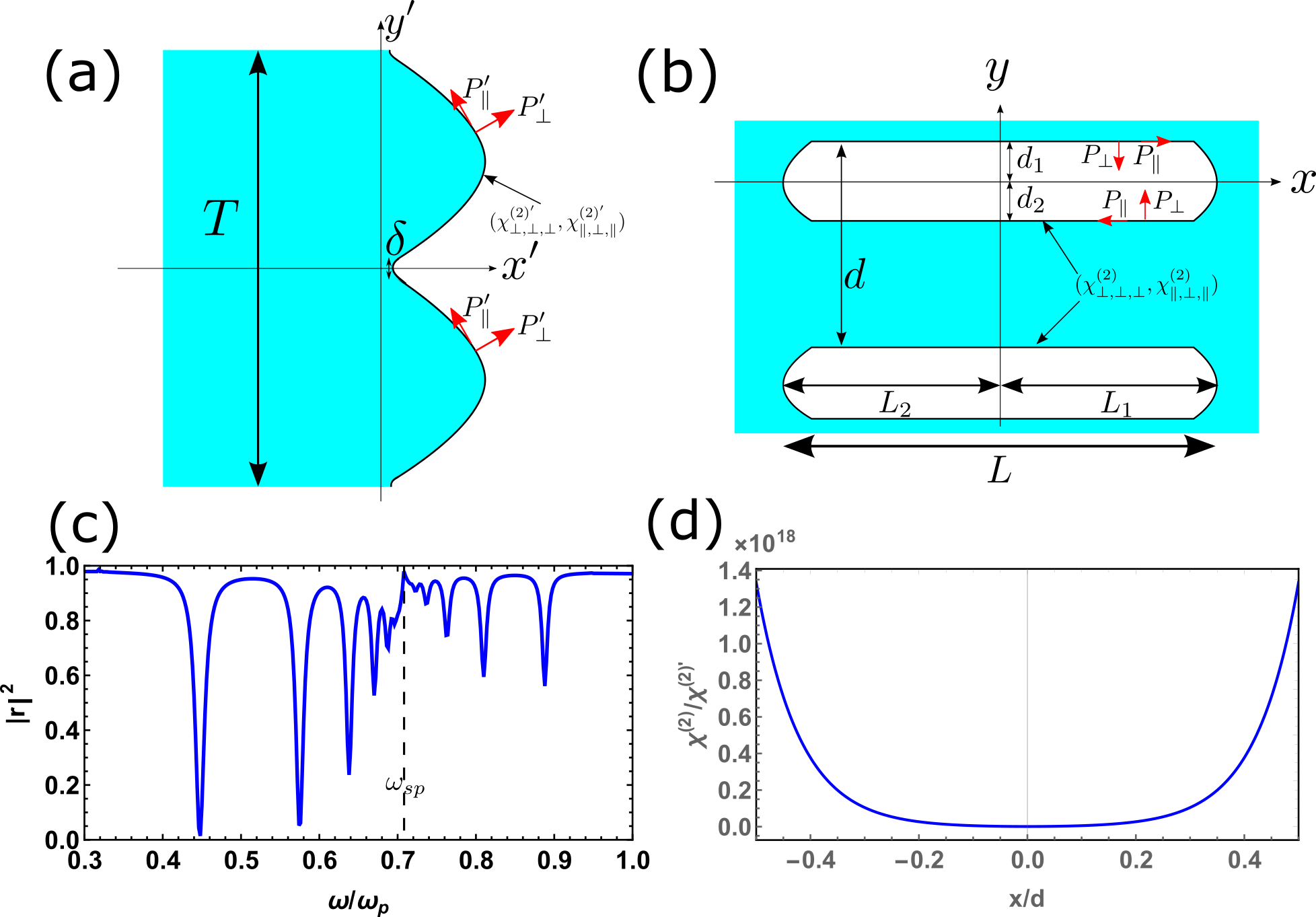}
\centering
\caption{Mapping of nonlinearity. (a) The geometry of the singular metasurface with the singularity scaled by $\delta$. The period is $T$. (b) The corresponding slab frame and the nonlinear surface polarization, where the slab period is $d$ and the cavity length is $L$. (c) The linear reflection spectrum of a singular metasurface, where the incident angle of plane wave is $\pi/4$. (d) The nonlinear surface susceptibility $\chi^{(2)}$ in the slab frame. The geometric parameter settings in panels (c) amd (d) are  $T=10$ nm, $\theta= 0.2\pi$, and $d_1=d_2=0.05d$.}
\label{Schematic}
\end{figure}

The singular metasurface investigated in this paper is shown in Fig. \ref{Schematic}(a). For a practical consideration, the sharp singular point is scaled by the width $\delta$. We consider the singular metasurface made up of a centrosymmetric plasmonic material parameterized with a Drude model $\varepsilon = 1- \omega_p^2/(\omega^2+ i \omega \gamma)$, where $\omega_p = 1.36 \times 10^{16}$ rad/s and $\gamma = 1\times 10^{14}$ rad/s. In a centrosymmetric medium, the second-order nonlinear process is not allowed \cite{boyd2020nonlinear}. However, the inversion symmetry can be broken at the material surface. It is reasonable then to define a surface second-order nonlinear susceptibility. We assume that all nonlinear contributions come from the metal surface \cite{reddy2019surface}. Following this model \cite{ciraci2012origin}, the two nonzero surface susceptibilities can be written as
\begin{equation}
\begin{split}
& \chi_{\bot \bot \bot}^{(2)'} = - \frac{\varepsilon_0}{4 n_0 e} \frac{3 \omega_F + i \gamma}{2 \omega_F + i \gamma} (\varepsilon_F-1)^2 \\
& \chi_{\parallel \bot \parallel}^{(2)'} = - \frac{\varepsilon_0}{2 n_0 e} (\varepsilon_F-1)^2
\end{split}
\label{surface susceptibity}
\end{equation}
where $\varepsilon_F$ is the relative permittivity at fundamental pump frequency $\omega_F$, $\varepsilon_0$ is the free space permittivity, $n_0 = 5.7\times 10^{28}$ m$^{-3}$ is the equilibrium charge density, and $-e$ is the electron charge \cite{ciraci2012origin}. The prime denotes the physical quantity in the metasurface frame, i.e. Fig. \ref{Schematic}(a). And the two surface polarizations can be expressed in the form
\begin{equation}
\begin{split}
& P_{\bot}' = \varepsilon_0 \chi_{\bot \bot \bot}^{(2)'} E_{\bot}^{'2} \\
& P_{\parallel}' = \varepsilon_0 \chi_{\parallel \bot \parallel}^{(2)'} E_{\bot}' E_{\parallel}'
\end{split}
\label{surface polarization}
\end{equation}
where the direction of $P_{\bot}'$ and $P_{\parallel}'$ is denoted in Fig. \ref{Schematic}(a), shown as red arrows. These two polarizations can be linked with two surface currents: the electric surface current $J^{e'}$ and the magnetic surface current $J^{m'}$. Their relations are \cite{albooyeh2015electromagnetic, reddy2017revisiting}
\begin{equation}
\begin{split}
& J^{m'} = \frac{1}{\varepsilon_b} \mathbf{n} \times \nabla_{\parallel} P_{\bot}' \\
& J^{e'} = \frac{\partial P_{\parallel}'}{\partial t} 
\end{split}
\end{equation}
where $\varepsilon_b$ is the background permittivity experienced by $P_{\bot}'$, and $J^{m'}$ and $J^{e'}$ are along the parallel direction. In our surface polarization model, the background permittivity experienced by the induced polarization at second-harmonic frequency $\omega_S$ is $\varepsilon_0 \varepsilon_S$ \cite{ciraci2012origin}, where $\varepsilon_S$ is the metal permittivity at frequency $\omega_S$.    

According to Eq. \eqref{surface polarization}, the fundamental field oscillating at frequency $\omega_F$ generates, through the second-order process, a nonlinear source that in turn gives rise to a pair of surface polarization components $(P_{\bot}',P_{\parallel}')$. These two polarizations lead to following boundary conditions at the second-harmonic frequency $\omega_S$ \cite{albooyeh2015electromagnetic, reddy2017revisiting}
\begin{equation}
\begin{split}
& E_{\parallel}^{'+} - E_{\parallel}^{'-} = - J_z^{m’} = - \frac{i k_{\parallel}'}{\varepsilon_b} P_{\bot}'  \\
& H_z^{'+} - H_z^{'-} = - J_{\parallel}^{e'} = i \omega_S P_{\parallel}'
\end{split}
\end{equation}
where the superscript + (-) represents the point immediately outside (inside) the metal interface.

Now we are going to map the above boundary conditions to the slab frame in Fig. \ref{Schematic}(b). A singular metasurface in Fig. \ref{Schematic}(a) with finite singularity can be mapped to a periodic slab system with the period $d$ in the $y$-direction and the finite cavity length $L$ shown in Fig. \ref{Schematic}(b) by following conformal mapping \cite{yang2018transformation}
\begin{equation}
\begin{split}
z = \frac{d}{2\pi} \mathrm{ln} \bigg( \frac{2}{e^{2\pi z'/T}-1)} + 1 \bigg)
\end{split}
\end{equation}
where $z=x + i y$ and $z'=x' + i y'$ are complex coordinates in the slab frame and the metasurface frame, respectively. When the size of singularity vanishes, the cavity length $L$ diverges, i.e. an infinite cavity. In this article, we set the parameters for slab period and cavity as $L=d=1$.  

From the rule of transformation optics, electric field $E_{\parallel}'$ and k-vector $k_{\parallel}'$ have been stretched by the same factor $1/\sqrt{\text{det}(\Lambda)}$ when mapping from the metasurface frame in Fig. \ref{Schematic}(a) to slab frame in Fig. \ref{Schematic}(b) \cite{pendry2006controlling, schurig2006calculation}. Here, $\mathrm{det}(\Lambda)$ is the Jacobian matrix whose elements are defined as $\Lambda_{i j} = \frac{\partial x_i'}{\partial x_j}$. Therefore, the surface polarization $P_{\bot}'$ is conserved from the metasurface frame to the slab frame. On the other hand, the z-component of magnetic field $H_z'$ is also conserved under the transformation, leaving the other surface polarization $P_{\parallel}$ unchanged as well. Since the surface polarization $P_{\bot,\parallel}'$ is a conserved quantity, we have    
\begin{equation}
\begin{split}
& P_{\bot}' = \varepsilon_0 \chi_{\bot \bot \bot}^{(2)'} E_{\bot}^{'2} = \varepsilon_0 \chi_{\bot \bot \bot}^{(2)'}  \frac{E_{\bot}^2}{\text{det}(\Lambda)}  = \varepsilon_0 \chi_{\bot \bot \bot}^{(2)} E_{\bot}^{2} = P_{\bot} \\
& P_{\parallel}' = \varepsilon_0 \chi_{\parallel \bot \parallel}^{(2)'} E_{\perp}' E_{\parallel}' = \varepsilon_0 \chi_{\parallel \bot \parallel}^{(2)'} \frac{E_{\bot} E_{\parallel}}{\text{det}(\Lambda)}  = \varepsilon_0 \chi_{\parallel \bot \parallel}^{(2)} E_{\bot} E_{\parallel} = P_{\parallel}  
\end{split}
\end{equation}
from which the transformation rule of the surface nonlinear susceptibility $\chi_{\bot \bot \bot,\parallel \bot \parallel}^{(2)}$ can be written as
\begin{equation}
\begin{split}
\chi_{\bot \bot \bot,\parallel \bot \parallel}^{(2)} &= \frac{\chi_{\bot \bot \bot,\parallel \bot \parallel}^{(2)'}}{\text{det}(\Lambda)}\\
 &= \chi_{\bot \bot \bot,\parallel \bot \parallel}^{(2)'} \left| \frac{d z}{d z'} \right|^2 \\
 &= \chi_{\bot \bot \bot,\parallel \bot \parallel}^{(2)'} \left( \frac{d}{T} \left| \sinh(\frac{2\pi}{d} z) \right| \right)^2
\end{split}
\end{equation} 
which is coordinate-dependent in the slab frame. In this frame, the metal-air interfaces are located at $y=d_1$ and $y=-d_2$. Therefore, the surface nonlinear susceptibility in the slab frame is a function of $x$, shown in Fig. \ref{Schematic}(d).

\section{Induced surface polarization}

\begin{figure}[ht]
\includegraphics[width=0.9\columnwidth]{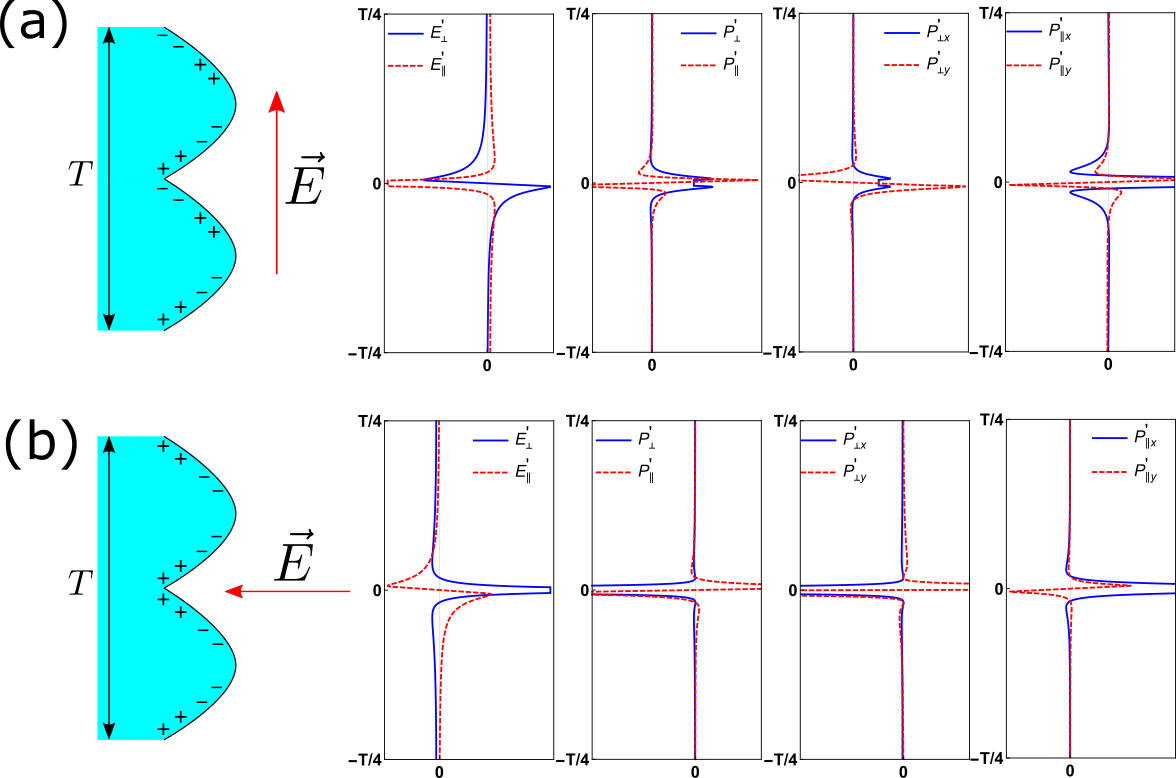}
\centering
\caption{Surface polarization of singular metasurface with period $T=10$ nm and vertex angle $\theta= 0.2\pi$. (a) $a_x$ mode excited by the parallel electric component at $\omega=0.449 \omega_p$ and the corresponding electric field and induced nonlinear surface polarization; (b) $a_y$ mode excited by the electric field normal to the metasurface at $\omega= 0.888 \omega_p$, and the corresponding electric field and induced nonlinear surface polarization.}
\label{SurfacePolarization}
\end{figure}

In above section, we have demonstrated that the surface polarization conserves under the transformation from the metasurface to the slab geometries. Thus, the nonlinear surface polarization profile $P_{\bot,\parallel}'(x',y')$ along the interface of metasurface can be obtained by directly mapping $P_{\bot,\parallel}(x,y)$ from slab frame. In Fig. \ref{SurfacePolarization}, the distribution of electric field at the interface $E_{\bot,\parallel}'$ and the induced polarization $P_{\bot,\parallel}'$ are shown. The singular metasurface supports two kinds of modes with different symmetries: the $a_x$ ($a_y$) mode below (above) $\omega_{sp}$ in the reflection spectrum in Fig. \ref{Schematic}(c) \cite{yang2018transformation}. In the reflection spectrum, the resonance peaks corresponds to the excitation of a few discrete modes, which merge into a continuous spectrum when the width of singularity $\delta \rightarrow 0$ \cite{yang2018transformation}. In Fig. \ref{SurfacePolarization}(a), we study the $a_x$ mode whose electric field is anti-symmetric for the normal field component $E_{\bot}'$ while symmetric for the tangential component $E_{\parallel}'$. As a comparison, the electric field of the $a_y$ mode shown in Fig. \ref{SurfacePolarization}(b) exhibits the opposite symmetry for the corresponding normal and tangential field components.

From the definition in Eq. \eqref{surface polarization}, the nonlinear surface polarization $P_{\bot}'$ is a quadratic function of electric field $E_{\bot}'$. Despite the different symmetry of the electric field $E_{\bot}'$ for $a_x$ and $a_y$ modes in Figs. \ref{SurfacePolarization}(a) and \ref{SurfacePolarization}(b), the surface polarization $P_{\bot}'$ of these two modes are both symmetric (an even function of $y'$). For the other surface polarization $P_{\parallel}'$, the results in Fig. \ref{SurfacePolarization} shows that both $a_x$ and $a_y$ modes have anti-symmetric property (an odd function of $y'$). This can be explained by observing that the electric fields $E_{\bot}'$ and $E_{\parallel}'$ share different symmetries for both $a_x$ and $a_y$ modes, so the product of these two fields ($\propto P_{\parallel}'$) remains anti-symmetric.       

In the right two columns in Fig. \ref{SurfacePolarization}, the two nonlinear surface polarizations $P_{\bot}'$ and $P_{\parallel}'$ are projected on $x$ and $y$ axes, respectively. We observe that the $x$-component is symmetric, while the y-component is anti-symmetric. In the following text, we show that the symmetry of these surface polarizations strongly affects the SHG in the far-field.

\section{Induced bulk polarization}

The induced nonlinear surface polarization becomes a source excitation at the second-harmonic frequency $\omega_S$. In order to provide an analytical solution for the excited second-harmonic field, we implement the calculation in the slab frame shown in Fig. \ref{Schematic}(b). The surface polarization at the metal-dielectric interface in the slab frame varies along the $x$-direction, which can be expanded as a Fourier series
\begin{equation}
\begin{split}
P_{\perp,\parallel} = \sum_{n=-\infty}^{\infty} P_{\perp,\parallel}^n e^{i k_n x}
\end{split} 
\end{equation}
With this mode expansion of the source excitation, the excited field can be also expanded as $H_z = \sum_{n=-\infty}^{\infty} H_z^n e^{i k_n x}$, where the mode $k_n$ in the slab frame can be expressed as
\begin{equation}
\begin{split}
    H_z^n(k_n,y)= \left\{
        \begin{array}{lr}
         b_+ e^{ - \sqrt{k_n^2} y} + b_- e^{\sqrt{k_n^2}y}, & -d_2 < y < d_1 \\
         c_+ e^{ - \sqrt{k_n^2} y} + c_- e^{\sqrt{k_n^2}y}, & -(d_2+d_3) < y < -d_2
        \end{array}
        \right.
\end{split} 
\end{equation}
and the corresponding tangential electric field is
\begin{equation}
\begin{split}
    E_x^n(k_n,y) =  \left\{
         \begin{array}{lr}
       - \frac{i \sqrt{k_n^2}}{\omega_S \varepsilon_0} \left( b_+ e^{ - \sqrt{k_n^2} y} - b_- e^{\sqrt{k_n^2}y} \right), & -d_2 < y < d_1 \\
      - \frac{i \sqrt{k_n^2}}{\omega_S \varepsilon_0 \varepsilon_S} \left( c_+ e^{ -\sqrt{k_x^2} y} - c_- e^{\sqrt{k_n^2}y} \right), &  -(d_2+d_3) < y < -d_2
        \end{array}
        \right.
\end{split}
\label{SH Ex}        
\end{equation}

\begin{figure}[ht]
\includegraphics[width=0.9\columnwidth]{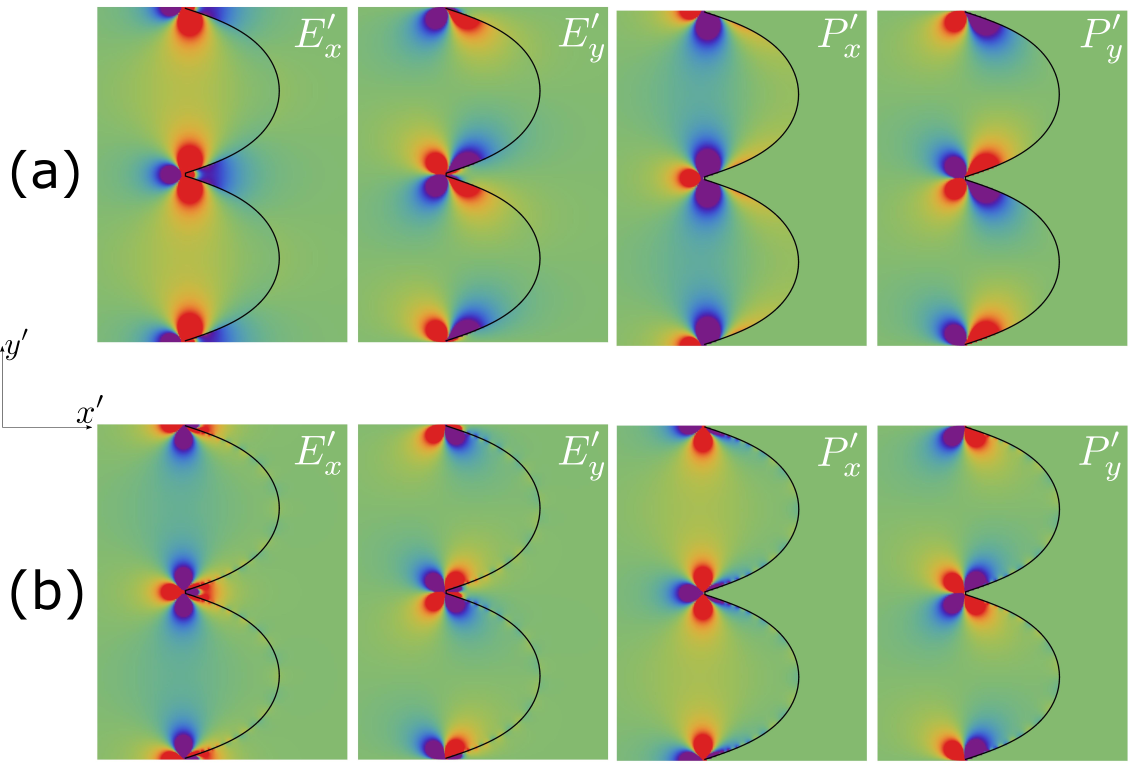}
\centering
\caption{Excited near field of singular metasurface with period $T=10$ nm and vertex angle $\theta= 0.2\pi$ in the second-harmonic frequency $\omega_S$. (a) Electric field and polarization profile for $a_x$ mode; (b) Electric field and polarization profile for $a_y$ mode.}
\label{NearField}
\end{figure}

The excited mode coefficients $b_{\pm}$ and $c_{\pm}$ can be obtained by imposing the following boundary conditions for second-harmonic near field
\begin{equation}
\begin{split}
& H_z^{n+} - H_z^{n-} \big|_{y=-d_2} = J_x^{e} \big|_{y=-d_2} = - i \omega_S P_{\parallel}^n \big|_{y=-d_2}  \\ 
& H_z^{n+} - H_z^{n-} \big|_{y=d_1} = -J_x^{e} \big|_{y=d_1} = i \omega_S P_{\parallel}^n \big|_{y=d_1} \\
& E_x^{n+} - E_x^{n-} \big|_{y=-d_2} = J_z^{m} \big|_{y=-d_2} = - \frac{i k_n}{\varepsilon_0 \varepsilon_S} P_{\bot}^n \big|_{y=-d_2} \\
& E_x^{n+} - E_x^{n-} \big|_{y=d_1} = -J_z^{m} \big|_{y=d_1} = - \frac{i k_n}{\varepsilon_0 \varepsilon_S} P_{\bot}^n \big|_{y=d_1}
\end{split}
\end{equation}
from which we obtain the following system of equation in the matrix form
{\footnotesize
\begin{equation}
\begin{split}
   \left(  \begin{array}{cccc}
     e^{|k_n|d_2} & e^{-|k_n|d_2} & -e^{|k_n|d_2} & -e^{-|k_n|d_2} \\
    e^{-|k_n|d_1} & e^{|k_n|d_1} & -e^{|k_n|(d_2+d_3)} & -e^{-|k_n|(d_2+d_3)} \\
    |k_n|e^{|k_n|d_2} & -|k_n|e^{-|k_n|d_2} & -|k_n|e^{|k_n|d_2}/\varepsilon_S & |k_n|e^{-|k_n|d_2}/\varepsilon_S \\
    |k_n|e^{-|k_n|d_1} & -|k_n|e^{|k_n|d_1} & -|k_n|e^{|k_n|(d_2+d_3)}/\varepsilon_S & |k_n|e^{-|k_n|(d_2+d_3)}/\varepsilon_S
    \end{array} \right)
    \left(  \begin{array}{c}
     b_+ \\ b_- \\ c_+ \\ c_-
     \end{array} \right)
  = 
     \left(  \begin{array}{c}
     - i \omega_S P_{\parallel}^n \big|_{y=-d_2} \\ i \omega_S P_{\parallel}^n \big|_{y=d_1} \\  \frac{\omega_S k_n}{\varepsilon_S} P_{\bot}^n \big|_{y=-d_2} \\ \frac{\omega_S k_n}{\varepsilon_S} P_{\bot}^n \big|_{y=d_1}
     \end{array} \right)
\end{split}        
\end{equation}
}

By solving the above system of equation, we obtain the excited near field by the surface polarization at $\omega_S$. In Figs. \ref{NearField}(a) and \ref{NearField}(b), we show the profile of near field and polarization for $a_x$ and $a_y$ modes, respectively. From Fig. \ref{SurfacePolarization} it can be observed that the nonlinear surface polarizations for $a_x$ and $a_y$ modes have the same kind of symmetry, thereby inducing the same symmetry for near field and polarization in Fig. \ref{NearField}.

\section{From near field to far field}

The induced surface polarization and near field at second-harmonic frequency are obtained in the previous section. The near field gives rise to an effective surface polarization macroscopically, which can be obtained by averaging the polarization in one period. Here, the grating period is assumed to be sub-wavelength so that the higher diffraction order can be ignored. In this approximation, the $k$-vector of second-harmonic field parallel to the metasurface is $k_{0y}^{\omega_S} = 2 k_{0y}^{\omega_F}$. This macroscopic surface polarization has two origins: one is from the surface polarization $P_{\bot,\parallel}'$ defined in Eq. \eqref{SurfacePolarization}, while the other one is from the bulk polarization induced by $P_{\bot,\parallel}'$. 

The application of transformation optics enables the calculation of these polarizations in the slab frame where the analytic calculation becomes possible. Let us first discuss the contribution from nonlinear surface polarization $P_{\bot,\parallel}'$, whose contribution to the effective surface polarization can be calculated by
\begin{equation}
\begin{split}
& \overline{P_{x1}'} = \frac{1}{T} \int P_x' d l' = \frac{1}{T} \left( \int P_{\perp}' d y' + \int P_{\parallel}' d x' \right) = \frac{1}{T} \left( \int P_{\perp} d y' + \int P_{\parallel} d x' \right) \\
& \overline{P_{y1}'} = \frac{1}{T} \int P_y' d l' = \frac{1}{T} \left( \int -P_{\perp}' d x' + \int P_{\parallel}' d y' \right) = \frac{1}{T} \left( \int -P_{\perp} d x' + \int P_{\parallel} d y' \right) 
\end{split}
\end{equation}
where the line integration is along the metasurface interface, and the quantity with overline stands for the effective surface polarization. Then, by using the chain rule and Cauchy-Riemann conditions, we have
\begin{equation}
\begin{split}
& \overline{P_{x1}'} = \frac{1}{T} \int \left( -P_{\perp} \frac{\partial x'}{\partial y} + P_{\parallel} \frac{\partial x'}{\partial x} \right) d x  \\
& \overline{P_{y1}'} = \frac{1}{T} \int \left( -P_{\perp} \frac{\partial x'}{\partial x} - P_{\parallel} \frac{\partial x'}{\partial y} \right) d x
\end{split}
\label{Effective surface polarization 1}
\end{equation}

The other contribution to the effective surface polarization is the induced bulk polarization in the metallic region at second-harmonic frequency $\omega_S$, which can be integrated as
\begin{equation}
\begin{split}
& \overline{P_{x2}'} = \frac{1}{T} \iint_{metal} P_x' dx' dy' \\
& \overline{P_{y2}'} = \frac{1}{T} \iint_{metal} P_y' dx' dy'
\end{split}
\end{equation} 

Since the transformation of bulk polarization follows \cite{luo2012transformation}
\begin{equation}
\begin{split}
    \left( {\begin{array}{c}
    P_x' \\
    P_y' 
    \end{array}} \right)
    = \frac{1}{\mathrm{det}(\Lambda)}
    \left( {\begin{array}{cc}
    \frac{\partial x'}{\partial x} & \frac{\partial x'}{\partial y} \\
    -\frac{\partial x'}{\partial y} & \frac{\partial x'}{\partial x} 
    \end{array}} \right)
    \left( {\begin{array}{c}
    P_x \\
    P_y 
    \end{array}} \right)
\end{split}
\end{equation}
and the bulk region integration follows \cite{arfken1999mathematical} 
\begin{equation}
\begin{split}
  \iint_{metal} dx' dy' = \iint_{metal} \mathrm{det}(\Lambda) dx dy
\end{split}
\end{equation}  
Therefore, we obtain the following integration formulas to calculate the contribution of bulk polarization to the effective surface polarization of the metasurface, which reads
\begin{equation}
\begin{split}
& \overline{P_{x2}'} = \frac{1}{T} \iint_{metal} \left( \frac{\partial x'}{\partial x} P_x + \frac{\partial x'}{\partial y} P_y \right) dx dy \\
& \overline{P_{y2}'} = \frac{1}{T} \iint_{metal} \left( -\frac{\partial x'}{\partial y} P_x + \frac{\partial x'}{\partial x} P_y \right) dx dy
\end{split}
\end{equation}  

The above two contributions can be summed up to form a total surface polarization
\begin{equation}
\begin{split}
& \overline{P_x'} = \overline{P_{x1}'} + \overline{P_{x2}'} \\
& \overline{P_y'} = \overline{P_{y1}'} + \overline{P_{y2}'}
\end{split}
\end{equation} 
from which an effective electric surface current and magnetic surface current can be obtained by
\begin{equation}
\begin{split}
& \overline{J_z^{m}} = \frac{i k_{0y}^{\omega_S}}{\varepsilon_0} \overline{P_x'} \\
& \overline{J_y^{e}} = - i \omega_S \overline{P_y'}
\end{split}
\label{Effective current}
\end{equation}  
Here, the background permittivity experienced by $\overline{P_x'}$ is $\varepsilon_0$.

We can now analyze how the symmetry of induced surface polarization $P_{\bot,\parallel}'$ and bulk polarization $P_{x,y}'$ affects the effective macroscopic surface polarization and current. The projection of $P_{\bot,\parallel}'$ along the $x$-direction is an even function (see Fig. \ref{SurfacePolarization}), whose average in a period gives a nonzero effective polarization $\overline{P_{x1}'} \neq 0$. On the other hand, the projection along the $y$-direction of the surface polarization $P_{\bot,\parallel}'$ is an odd function, resulting in a zero effective polarization contribution, $\overline{P_{y1}'}=0$. Similarly, the bulk polarization in Fig. \ref{NearField} also demonstrates that $P_x'$ is even while $P_y'$ is odd, resulting in $\overline{P_{x2}'} \neq 0$ and $\overline{P_{y2}'} = 0$. Therefore, the total effective surface polarization is nonzero for $\overline{P_x'}$ but zero for $\overline{P_y'}$. Then from the relation between surface current and polarization in Eq. \eqref{Effective current}, we conclude that second-order nonlinear effect in singular metasurface gives a nonzero effective magnetic surface current but a zero electric current.   

\begin{figure}[ht]
\includegraphics[width=0.9\columnwidth]{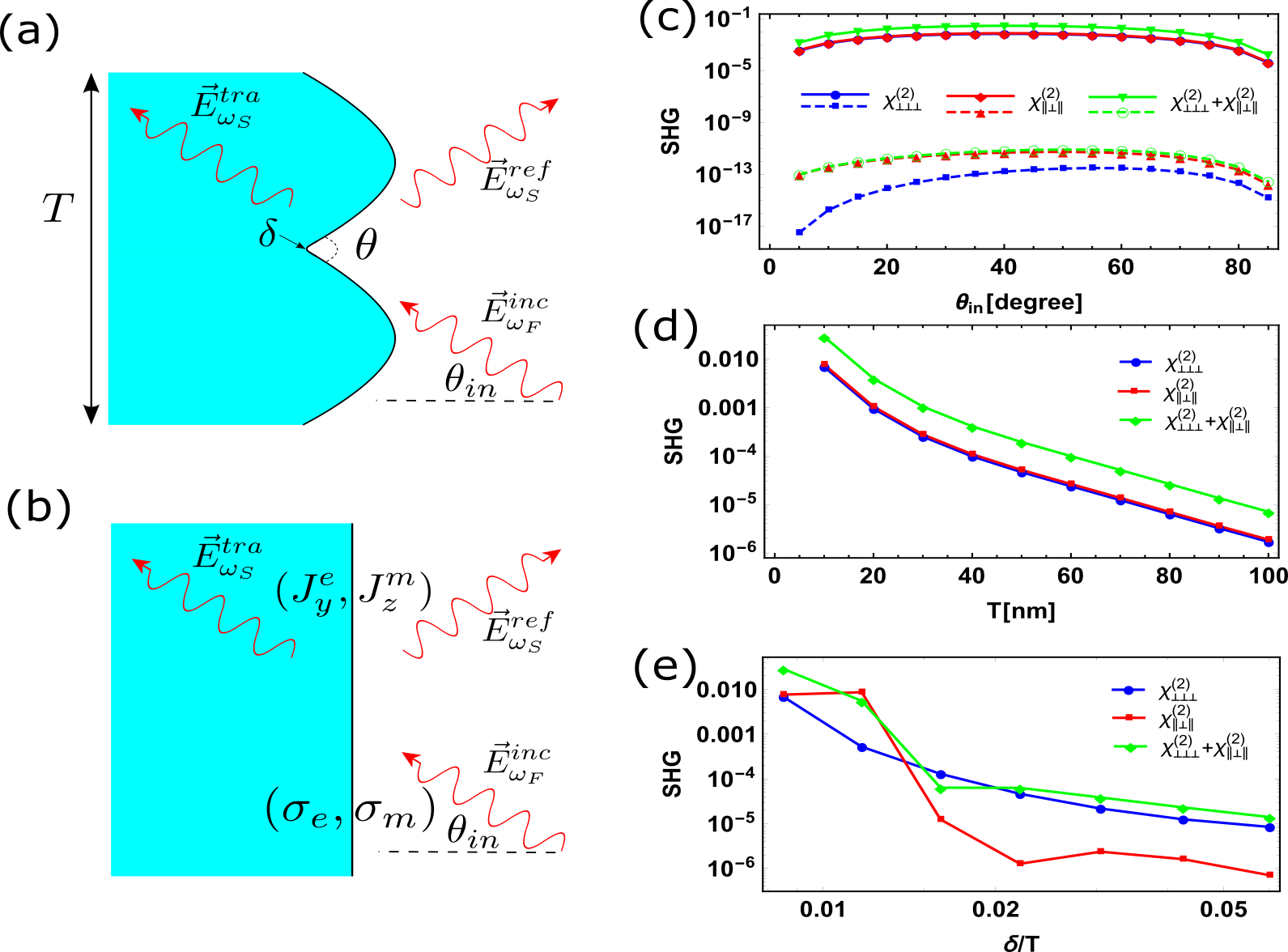}
\centering
\caption{Far field calculation for the singular metasurface. (a) The pump field at fundamental frequency $\omega_F$ excites SHG in the far field at $\omega_S$; (b) Simplified flat surface geometry for SHG calculation, where the nonlinear source is modeled as two current source $(J_y^e, J_z^m)$; (c) The dependence of SHG on incident angle of plane wave $\theta_{in}$ for singular metasurface with period $T=10$ nm; (d) The dependence of SHG on period of metasurface $T$; (e) The dependence of SHG on size of singularity $\delta$ with period $T=10$ nm. The vertex angle of singular metasurface is fixed as $\theta= 0.2\pi$.}
\label{FarField}
\end{figure}

The effective surface currents generate a second-harmonic wave in the far field, shown in Fig. \ref{FarField}(a). Assuming the grating period is sub-wavelength, the metasurface at $\omega_S$ can be simplified as a flat metal surface with a pair of surface current $(J_y^e, J_z^m)$ and a pair of surface conductivity $(\sigma_e,\sigma_m)$ (electric and magnetic), see Fig. \ref{FarField}(b), where the calculation of two surface conductivities is given in linear response theory of singular metasurface \cite{yang2018transformation}. The generated far field is expressed as
\begin{equation}
\begin{split}
& H_z^{\omega_S} =
\left\{ {\begin{array}{lr}
    r^{\omega_S} H_0 e^{i k_{0x}^{\omega_S} x + i k_{0y}^{\omega_S}}, & x>0 \\
    t^{\omega_S} H_0 e^{-i k_{0x}^{'\omega_S} x + i k_{0y}^{\omega_S}}, & x<0 
    \end{array}} \right.
\end{split}
\end{equation}
where $r^{\omega_S}$ and $t^{\omega_S}$ are defined as the coefficients for reflected and transmitted second-harmonic fields, respectively. Since the metasurface studied in this paper is semi-infinite, we only consider reflected second-harmonic field, whose generation efficiency is $|r^{\omega_S}|^2$. By matching the tangential field $H_z^{\omega_S}$ and $E_y^{\omega_S}$ at the interface $x=0$ in Fig. \ref{FarField}(b), we arrive at
\begin{equation}
\begin{split}
 \left( {\begin{array}{cc}
  1 & -1 \\
  Z_a & Z_m
 \end{array}} \right)
  \left( {\begin{array}{c}
  r^{\omega_S} H_0 \\
  t^{\omega_S} H_0
 \end{array}} \right)
 =
 -\left( {\begin{array}{c}
  J_y^e \\
  J_z^m
 \end{array}} \right)
 -\left( {\begin{array}{cc}
  \frac{\sigma_e Z_a}{2} & -\frac{\sigma_e Z_m}{2} \\
  \frac{\sigma_m}{2} & \frac{\sigma_m}{2}
 \end{array}} \right)
  \left( {\begin{array}{c}
  r^{\omega_S} H_0 \\
  t^{\omega_S} H_0
 \end{array}} \right)
\end{split}
\end{equation}
where $Z_a = \frac{k_{0x}^{\omega_S}}{\omega_S \varepsilon_0}$ and $Z_m = \frac{k_{0x}^{'\omega_S}}{\omega_S \varepsilon_0 \varepsilon_S}$. After some algebraic manipulation, we obtain
\begin{equation}
\begin{split}
  \left( {\begin{array}{c}
  r^{\omega_S} H_0 \\
  t^{\omega_S} H_0
 \end{array}} \right)
 =
 -
 \left( {\begin{array}{cc}
  1+\frac{\sigma_e Z_a}{2} & -(1+\frac{\sigma_e Z_m}{2}) \\
  Z_a + \frac{\sigma_m}{2} & Z_m+\frac{\sigma_n}{2}
 \end{array}} \right)^{-1}
 \left( {\begin{array}{c}
  J_y^e \\
  J_z^m
 \end{array}} \right)
\end{split}
\end{equation}
and the SHG efficiency is 
\begin{equation}
\begin{split}
  |r^{\omega_S}|^2 = \left| \frac{2(J_z^m(2+Z_m \sigma_e) + J_y^e (2 Z_m + \sigma_m))}{4 (Z_a + Z_m + Z_a Z_m \sigma_e)  + (4+(Z_a+Z_m)\sigma_e)\sigma_m}  \right|^2 / \left|H_0\right|^2
\end{split}
\label{SHG singular metasurface}
\end{equation}

In Fig. \ref{FarField}(c), we calculate the SHG efficiency as a function of the incident angle for an incident plane wave of frequency $\omega_F = 0.449 \omega_p$, the first resonance peak in Fig. \ref{Schematic}(c). The pump field is a TM wave with a peak intensity of 55 MW/cm$^2$. By changing the incidence angle, it is surprising to observe that the SHG shows a quite flat curve (solid lines), which means the SHG of the singular metasurface weakly depends on $\theta_{in}$. Therefore, the singular metasurface can realize all-angle SHG. This weak dependence on the incident angle is due to a hidden dimension of the singular metasurface \cite{pendry2017compacted}. The additional dimension gives the $k$-vector of the mode one more degree of freedom, making the mode excitation in a continuous manner when changing the incident angle. We have also divided the SHG contribution from the two components of nonlinear surface susceptibility $\chi_{\bot \bot \bot}^{(2)}$ and $\chi_{\parallel \bot \parallel}^{(2)}$. The blue curve corresponds to the case when only $\chi_{\bot \bot \bot}^{(2)}$ is considered, while the red curve refers to the contribution of $\chi_{\parallel \bot \parallel}^{(2)}$. The green curve shows the case when both surface susceptibilities are considered. These theoretical results have also been confirmed with numerical simulation, where the detailed simulation setup and results can be found in Appendixes \ref{AppendixA} and \ref{AppendixC}. 

For reference, we compare the SHG from the singular metasurface with the SHG from an unstructured metal surface, i.e., a flat surface. The SHG from a flat surface made up of the same metal as the metasurface is presented in Fig. \ref{FarField}(c) as dashed lines, where three different colors correspond to three kinds of combinations of nonlinear surface susceptibilities. The comparison in Fig. \ref{FarField}(c) proves the singular metasurface strongly improves the SHG efficiency by ten orders of magnitude when compared with a flat surface. Details about the calculation of a flat metal surface can be found in Appendix \ref{AppendixB}. 

In Fig. \ref{FarField}(d), we keep the incident angle at 45 degrees but change the grating period $T$ in the calculation of SHG. In tuning the parameter $T$, the shape of the metasurface is preserved. These results show that the SHG from singular metasurface decreases when increasing the grating period. This is expected because the number of singular points reduces in a given area when $T$ increases, leading to weaker harmonic generation. 

In above far-field calculation of SHG, the shape of the singular metasurface preserves such that the ratio of singularity size $\delta$ to period $T$ is a constant. The ratio $\delta/T$ characterizes the degree of singularity, where $\delta/T \rightarrow 0$ gives an ideal singular point. In Fig. \ref{FarField}(e), we study how $\delta/T$ affects the SHG in the far field by fixing the grating period $T$ and simultaneously shrinking the size of singular point $\delta$, which corresponds to elongating the cavity length $L$ in the slab frame. The calculation results show that the SHG increases when reducing $\delta/T$ and finally diverges as $\delta/T \rightarrow 0$. However, both the nonclassical effects from electrons \cite{yang2019nonlocal} and experimental imperfection \cite{yang2018transformation} result in a blunt singular point for surface plasmons, which ends up with a finite SHG. Another takeaway emerging from Figs. \ref{FarField}(d) and \ref{FarField}(e) is that the SHG efficiency is determined by the size of the singular point and grating period it is possible to achieve.

\section{SHG at normal incidence}

As shown in the previous section, the nonlinear source in the singular metasurface can be modeled as two effective surface currents $(J_y^e, J_z^m)$ with only the magnetic one being nonzero. From the definition in Eq. \eqref{Effective current}, the effective magnetic surface current is proportional to wave vector $k_{0y}^{\omega_S}$. Therefore, both electric and magnetic surface currents become zero under normal incidence, resulting in a zero harmonic field generation in the far field. This situation is often impractical in experimental setups, where instead being able to generate at normal incidence is preferable. Is it then possible to modify the metasurface in order to generate SHG in the far-field under normal incidence? 

\begin{figure}[ht]
\includegraphics[width=0.9\columnwidth]{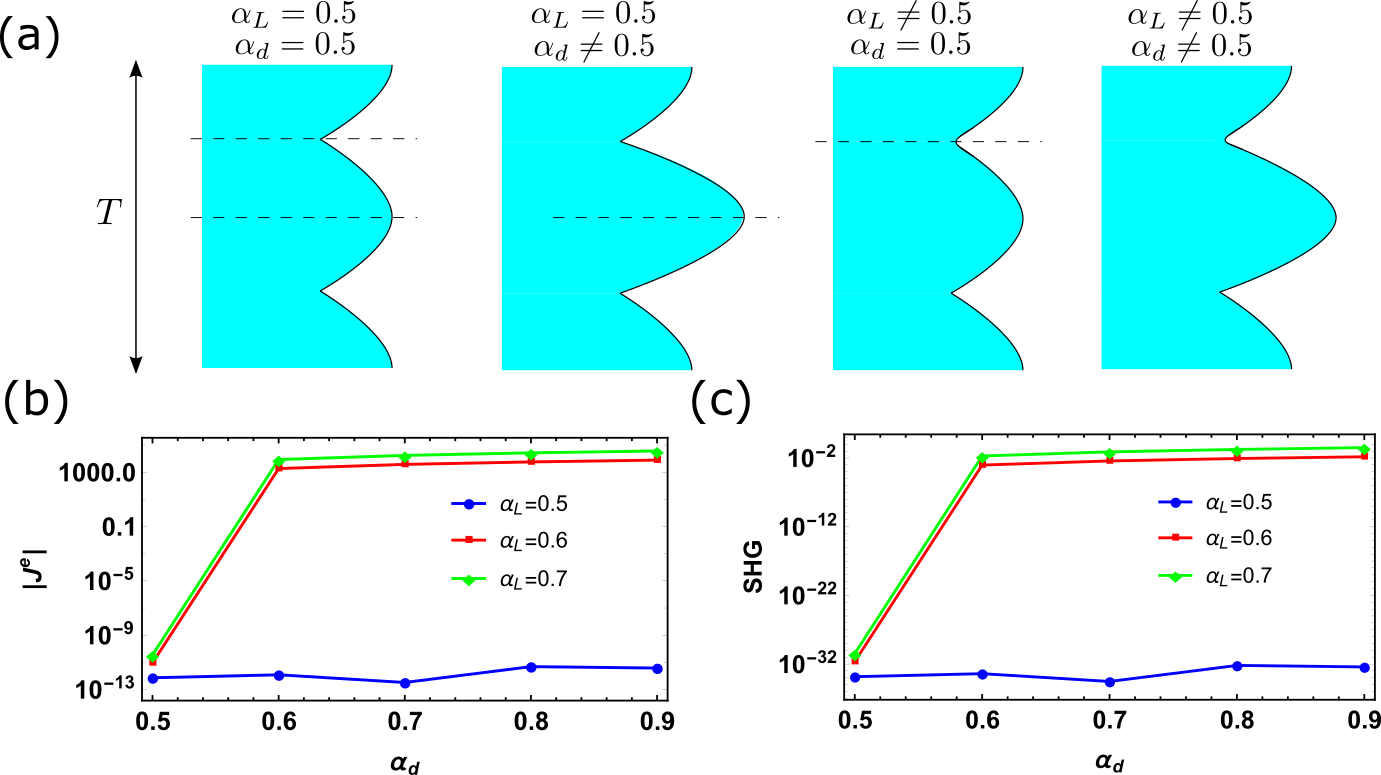}
\centering
\caption{SHG under normal incidence by breaking the surface inversion symmetry. (a) symmetry of singular metasurface quantified by two parameter $\alpha_L$ and $\alpha_d$, where dashed line show the inversion center; (b) The effective electric surface current as a function of $\alpha_L$ and $\alpha_d$; (c) SHG as a function of $\alpha_L$ and $\alpha_d$. The parameter setting for the singular metasurfaces are $T=10$ nm and $\theta= 0.2\pi$.}
\label{FarField_Asymmetric}
\end{figure}

The answer is yes. The singular metasurface shown in Fig. \ref{Schematic}(a) possesses two kinds of inversion symmetries along with the interface. To quantify the degree of asymmetry for the singular metasurface, we define two factors: $\alpha_L$ and $\alpha_d$. $\alpha_L$ is defined as $L_1/L$ in Fig. \ref{Schematic}(b), quantifying the asymmetry of two singular points in one period. In contrast, $\alpha_d$ measures the asymmetry of two bumps of singular surface, defined as $\alpha_d = d_1/(d_1+d_2)$ in Fig. \ref{Schematic}(b). Particularly, $\alpha_L=0.5$ gives two identical singular points, while $\alpha_d=0.5$ leads to two bumps with the same shape. In Fig. \ref{FarField_Asymmetric}(a), we have shown four cases of the singular metasurface: (1) $\alpha_L=0.5$ and $\alpha_d=0.5$; (2) $\alpha_L=0.5$ and $\alpha_d \neq 0.5$; (3) $\alpha_L \neq 0.5$ and $\alpha_d=0.5$; (4) $\alpha_L \neq 0.5$ and $\alpha_d \neq 0.5$. For the pure symmetric case (1), two kinds of inversion centers are marked as black dashed lines. For the asymmetric bump shape in case (2), the inversion center is the center of the bumped region. In contrast, for asymmetric singular points in case (3), the inversion center locates at the center of the singularity. Finally, when the symmetries of both the pumped region and singular points are broken, the surface inversion center disappears. Therefore, we expect only case (4) without a surface inversion symmetry to contribute an SHG in the far field.

To confirm the above assertion, we calculate the effective surface current and SHG in the parameter space of $\alpha_L$ and $\alpha_d$. Note that changing the value of $\alpha_L$ and $\alpha_d$ shifts the resonance peak position, so we calculate SHG for the corresponding first-order mode. Despite a lack of inversion symmetry, the effective magnetic surface current keeps zero under normal incidence. However, the story differs for the effective electric surface current $\overline{J^e}$. Fig. \ref{FarField_Asymmetric}(b) illustrates the evolution of $\overline{J^e}$ in parameter space, which shows the nonzero effective electric surface can only be achieved by deviating both $\alpha_L$ and $\alpha_d$ from 0.5. The nonzero surface electric current subsequently gives rise to a large SHG in the far field, shown in Fig. \ref{FarField_Asymmetric}(c).

\section{Conclusion}

In conclusion, we have studied the SHG for singular metasurfaces following a transformation optics approach. By means of transformation optics we have mapped a complex singular surface into a simple slab geometry, such that an analytical solution to SHG can be obtained. Also, the nonlinear source of SHG is modeled as a surface polarization, whose symmetry determines the induced near and far-field pattern. In addition, the singular metasurface in this paper possesses a variety of symmetries. We show that broken symmetry is necessary to receive a nonlinear far-field signal, which can be realized by using oblique incidence or by breaking the surface inversion symmetry. Finally, we found that SHG of the singular metasurface weakly depends on the incident angle of light, which can be applied to realize an all-angle SHG device.

\section*{Acknowledgments}
The authors thank Y. Sivan for useful discussions. F.Y. acknowledges the Fundamental Research Funds for the Central Universities and Science Specialty Program (Grand . No. 2020SCUNL210) from Sichuan University. 

\appendix

\section{Comsol setup for SHG calculation}
\label{AppendixA}
To check the validity of our theory, we have implemented a Comsol model to perform numerical simulations. We solve a system of two one-way coupled equations for the fundamental and second-harmonic fields, respectively. The critical point in the numerical simulations is the setup of the two surface polarizations $P_{\perp}$ and $P_{\parallel}$ in Comsol. Accounting for $P_{\parallel}$ is straightforward as it is directly related to an electric surface current by $J_{\parallel}^e = - i \omega_S P_{\parallel}$, which can be easily implemented using Comsol built-in source options.
Considering a surface polarization normal to the interface, on the other hand, is not as simple.
In order to account for such a polarization we need to add a weak form contribution.
For a bulk polarization vector, ${\bf P}_b$, this is $\mu_0\omega^2_S\int_\Omega{\bf P}_b\cdot\tilde{\bf E}dV$ where  $\tilde{\bf E}$ is the test function and $\Omega$ is the volume inside the metal. 
The relation between a bulk polarization and the surface polarization, ${\bf P}$, can be expressed as ${\bf P}_b=\delta_{\partial\Omega}{\bf P}$, where the delta function $\delta_{\partial\Omega}$ is non-zero only at the metal surface.
Finally, for a surface polarization normal to the surface, i.e. ${\bf P}=P_\perp\hat{\bf n}$, we get the following weak form contribution:
\begin{equation}
\begin{split}
\mu_0\omega^2_S\int_{\partial\Omega} P_{\perp}\tilde{\bf E}\cdot\hat{\bf n}dS
\end{split}
\end{equation}
where the integral is now performed only on the metal boundary $\partial\Omega$.
Both $P_{\perp}$ and $\tilde{\bf E}$ are evaluated inside the metal using the built-in operators \textit{down} or \textit{up}.
With this setup in Comsol, we were able to perform numerical simulations of SHG from singular metasurfaces.

\section{SHG from a flat surface}
\label{AppendixB}
Now, we assume a TM-polarized plane wave ($E_x, E_y, H_z$) incident on a flat metallic surface ($x=0$) from air with incident angle $\theta_{in}$, where $x>0$ and $x<0$ correspond to air and metal domain, respectively. The transmitted field inside the metal can be expressed as $H_z^{tra} = t H_0 e^{- i k_{0x}^{'\omega_F} x + i k_{0y}^{\omega_F} y}$ with $k_{0y}^{\omega_F} = k_{0}^{\omega_F} \sin \theta_{in}$, where $t$ is transmission coefficient for magnetic field and can be written in terms of incident angle as 
\begin{equation}
\begin{split}
t = \frac{2 \varepsilon_F \cos \theta_{in}}{\varepsilon_F \cos \theta_{in} + \sqrt{\varepsilon_F - \sin^2 \theta_{in}}}
\end{split}
\end{equation}

The electric field at $\omega_F$ inside the metal ($x<0$) is 
\begin{equation}
\begin{split}
E_x = - \frac{k_{0y}^{\omega_F}}{\omega_F \varepsilon_0 \varepsilon_F} t H_0  e^{- i k_{0x}^{'\omega_F} x + i k_{0y}^{\omega_F} y} \\
E_y = - \frac{k_{0x}^{'\omega_F}}{\omega_F \varepsilon_0 \varepsilon_F} t H_0  e^{- i k_{0x}^{'\omega_F} x + i k_{0y}^{\omega_F} y}
\end{split}
\end{equation}
From the electric field, the surface polarization can be obtained as
\begin{equation}
\begin{split}
P_x = \varepsilon_0 \chi_{\bot \bot \bot}^{(2)} (E_{x})^2 = \chi_{\bot \bot \bot}^{(2)} \frac{(k_{0y}^{\omega_F})^2}{\omega_F^2 \varepsilon_0 \varepsilon_F^2} t^2 H_0^2 e^{i k_{0y}^{\omega_S} y} \\
P_y =  \varepsilon_0 \chi_{\parallel \bot \parallel}^{(2)} E_x E_y  = \chi_{\parallel \bot \parallel}^{(2)} \frac{k_{0x}^{'\omega_F} k_{0y}^{\omega_F}}{\omega_F^2 \varepsilon_0 \varepsilon_F^2} t^2 H_0^2 e^{i k_{0y}^{\omega_S} y}
\end{split}
\end{equation}
where $k_{0y}^{\omega_S} = 2 k_{0y}^{\omega_F} $. Finally, the two surface currents can be expressed as
\begin{equation}
\begin{split}
J_z^m = i \frac{k_{0y}^{\omega_S}}{\varepsilon_0 \varepsilon_S} \chi_{\bot \bot \bot}^{(2)} \frac{(k_{0y}^{\omega_F})^2}{\omega_F^2 \varepsilon_0 \varepsilon_F^2} t^2 H_0^2 e^{i k_{0y}^{\omega_S} y}\\
J_y^e =  - i \omega_S \chi_{\parallel \bot \parallel}^{(2)} \frac{k_{0x}^{'\omega_F} k_{0y}^{\omega_F}}{\omega_F^2 \varepsilon_0 \varepsilon_F^2} t^2 H_0^2 e^{i k_{0y}^{\omega_S} y}
\end{split}
\end{equation}

The generated second-harmonic field can be easily obtained with these two currents by mode matching. The generated second-harmonic field by the surface current $(J^e, J^m)$ are expressed as 
\begin{equation}
\begin{split}
H_z^{ref(\omega_S)} &= r^{\omega_S} H_0 e^{i k_{0x}^{\omega_S} x + i k_{0y}^{\omega_S} y} \\
H_z^{tra(\omega_S)} &= t^{\omega_S} H_0 e^{-i k_{0x}^{'\omega_S} x + i k_{0y}^{\omega_S} y}
\end{split}
\end{equation}
where $r^{\omega_S}$ and $t^{\omega_S}$ are the corresponding coefficient for reflected and transmitted second-harmonic fields. Then by matching the field at the boundary with following boundary conditions
\begin{equation}
\begin{split}
H_z^{\omega_S+} - H_z^{\omega_S-} &= - J_y^e \\
E_y^{\omega_S+} - E_y^{\omega_S-} &= - J_z^m
\end{split}
\end{equation}
we have the following SHG efficiency for the flat metal surface 
\begin{equation}
\begin{split}
|r^{\omega_S}|^2 =  \left| \frac{Z_m J_y^e + J_z^m}{Z_m + Z_d}   \right|^2  / \left|H_0\right|^2
\end{split}
\label{SHG flat surface}
\end{equation}

\begin{figure}[ht]
\includegraphics[width=0.4\columnwidth]{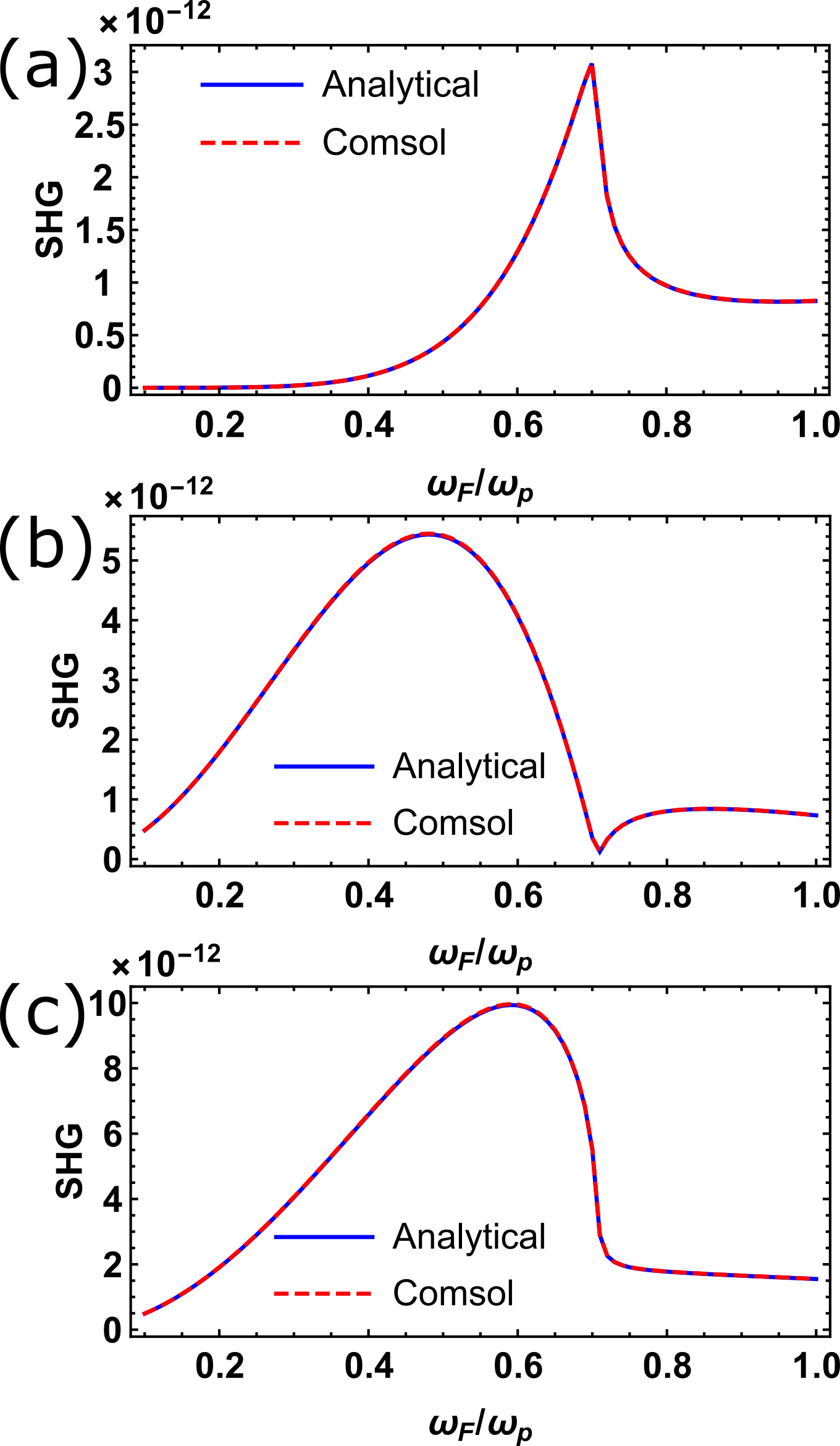}
\centering
\caption{SHG from a flat metallic surface. (a) $\chi_{\bot \bot \bot}^{(2)} \neq 0$ and $\chi_{\parallel \bot \parallel}^{(2)} = 0$; (b) $\chi_{\bot \bot \bot}^{(2)} = 0$ and $\chi_{\parallel \bot \parallel}^{(2)} \neq 0$; (c) $\chi_{\bot \bot \bot}^{(2)} \neq 0$ and $\chi_{\parallel \bot \parallel}^{(2)} \neq 0$. The solid line and dashed line correspond to theoretical calculation with Eq. \ref{SHG flat surface} and Comsol simulation, respectively.}
\label{Flatsurface_SHG}
\end{figure}

As a benchmark, we check the theoretical calculation of SHG efficiency of a flat metal surface by Eq. \ref{SHG flat surface} with Comsol simulation in Fig. \ref{Flatsurface_SHG}, where three combinations of two surface susceptibility are considered. The excellent agreement between theory and simulation demonstrates the correctness of our theoretical approach.

\section{Numerical verification}
\label{AppendixC}
\begin{figure}[ht]
\includegraphics[width=0.4\columnwidth]{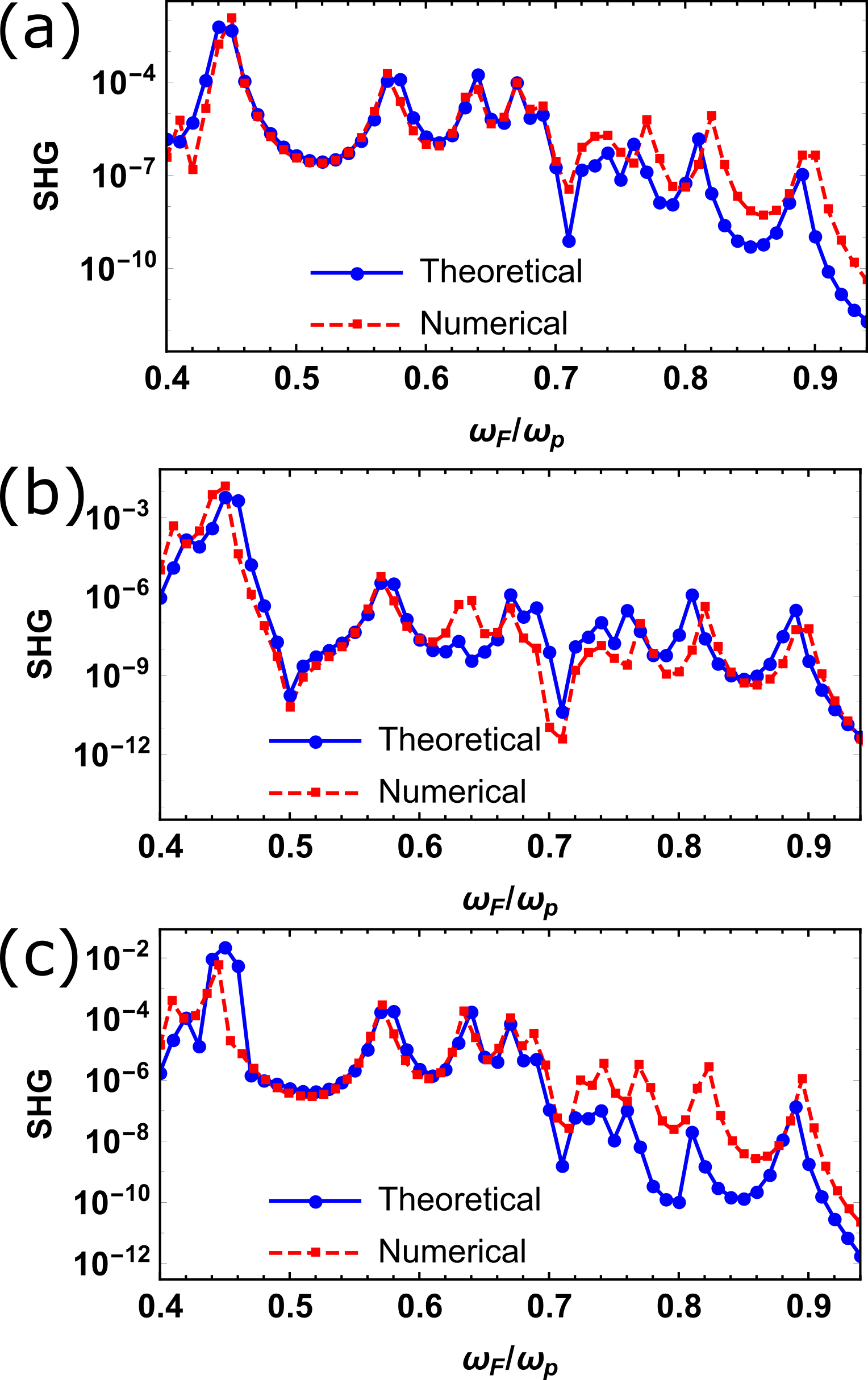}
\centering
\caption{Numerical verification for theoretical calculation of SHG from singular metasurfaces. (a) $\chi_{\bot \bot \bot}^{(2)} \neq 0$ and $\chi_{\parallel \bot \parallel}^{(2)} = 0$; (b) $\chi_{\bot \bot \bot}^{(2)} = 0$ and $\chi_{\parallel \bot \parallel}^{(2)} \neq 0$; (c) $\chi_{\bot \bot \bot}^{(2)} \neq 0$ and $\chi_{\parallel \bot \parallel}^{(2)} \neq 0$. The solid line and dashed line correspond to theoretical calculation with Eq. \ref{SHG singular metasurface} and Comsol simulation, respectively.}
\label{Numerical_Rigorous}
\end{figure} 

Using the Comsol simulation, we further check the correctness of our theoretical calculations. In Fig. \ref{Numerical_Rigorous}, we compare our theory with Comsol results, where a good agreement is achieved. A small discrepancy comes from neglecting the integration of the polarization in the region $x>L_1$ and $x<-L_2$ in the slab frame.

\bibliography{reference}

\end{document}